\documentclass{article}
\usepackage{PRIMEarxiv}
\usepackage[utf8]{inputenc} % allow utf-8 input
\usepackage[T1]{fontenc}    % use 8-bit T1 fonts
\usepackage{hyperref}       % hyperlinks
\usepackage{url}            % simple URL typesetting
\usepackage{booktabs}       % professional-quality tables
\usepackage{amsfonts}       % blackboard math symbols
\usepackage{nicefrac}       % compact symbols for 1/2, etc.
\usepackage{microtype}      % microtypography
\usepackage{lipsum}
\usepackage{fancyhdr}       % header
\usepackage{graphicx}       % graphics
\graphicspath{{media/}}     % organize your images and other figures under media/ folder
\usepackage{lscape}
\usepackage{tabularx}
\usepackage{xcolor,colortbl}
\usepackage{caption}
\title{An overview of open source
Deep Learning-based libraries for
Neuroscience
}

\author{
  Louis Fabrice Tshimanga \\
  Department of Neuroscience (DNS) \\
  University of Padova \\
  \texttt{louisfabrice.tshimanga@unipd.it} \\
   \And
  Manfredo Atzori \\
  Department of Neuroscience (DNS), \\
  Padova Neuroscience Center (PNC) \\
  University of Padova \\
  Information Systems Institute \\
  University of Applied Sciences Western Switzerland (HES-SO Valais)\\
  \texttt{manfredo.atzori@unipd.it} \\
  \And
  Federico Del Pup \\
  Department of Neuroscience (DNS), \\
  Department of Information Engineering (DEI) \\
  University of Padova \\
  \texttt{federico.delpup@studenti.unipd.it} \\
  \And
  Maurizio Corbetta \\
  Department of Neuroscience (DNS), \\
  Padova Neuroscience Center (PNC) \\
  University of Padova \\
  Department of Neurology \\
  Washington University School of Medicine \\
  \texttt{maurizio.corbetta@unipd.it} \\
}

\begin{document}
\maketitle

\begin{abstract}
%Background
In recent years, deep learning revolutionized machine learning and its applications, producing results comparable to human experts in several domains, including neuroscience.
Each year, hundreds of scientific publications present applications of deep neural networks for biomedical data analysis.
%motivation
Due to the fast growth of the domain, it could be a complicated and extremely time-consuming task for worldwide researchers to have a clear perspective of the most recent and advanced software libraries.
%objective
This work contributes to clarify the current situation in the domain, outlining the most useful libraries that implement and facilitate deep learning application to neuroscience, allowing scientists to identify the most suitable options for their research or clinical projects.
%Methods
This paper summarizes the main developments in Deep Learning and their relevance to Neuroscience;
it then reviews neuroinformatic toolboxes and libraries, collected from the literature and from specific hubs of software projects oriented to neuroscience research.
The selected tools are presented in tables detailing key features grouped by domain of application (e.g. data type, neuroscience area, task), model engineering (e.g. programming language, model customization) and technological aspect (e.g. interface, code source).
%Results
The results show that, among a high number of available software tools, several libraries are standing out in terms of functionalities for neuroscience applications.
%Discussion
The aggregation and discussion of this information can help the neuroscience community to devolop their research projects more efficiently and quickly, both by means of readily available tools, and by knowing which modules may be improved, connected or added.
\newline
%\textbf{Keywords:} Deep Learning, Neuroscience, Open-source

\end{abstract}

% keywords can be removed

\section{Introduction}
In the last decade, Deep Learning (DL) has taken over most classic approaches in Machine Learning (ML), Computer Vision, Natural Language Processing, showing an unprecedented versatility, and matching or surpassing the performances of human experts in narrow tasks.\\
The recent growth of DL applications to several domains, including Neuroscience, consequently offers numerous open-source software opportunities for researchers.  \\
Mapping available resources can allow a faster and more precise exploitation.\\
Neuroscience is a diversified field on its own, as much for the objects and scales it focuses on, as for the types of data it relies on. \\
The discipline is also historically tied to developments in electrical, electronic, and information technology. 
Modern Neuroscience relies on computerization in many aspects of data generation, acquisition, and analysis. 
Statistical and Machine Learning techniques already empower many software packages, that have become de facto standards in several subfields of Neuroscience, such as Principal and Independent Component Analysis in Electroencephalography and Neuroimaging, to name a few.\\
Meanwhile, the rich and rapidly evolving taxonomy of Deep Neural Networks (DNNs) is becoming both an opportunity and hindrance.
On the one hand, currently open-source DL libraries allow an increasing number of applications and studies in Neuroscience.
On the other hand, the adoption of available methods is slowed down by a lack of standards, reference frameworks and established workflows. 
Scientific communities whose primary focus or background is not in machine learning engineering may be left partially aside from the ongoing Artificial Intelligence (AI) gold rush.
\\
For such reasons it is fundamental to overview open-source libraries and toolkits. Framing a panorama could help researchers in selecting ready-made tools and solutions when convenient, as well as in pointing out and filling in the blanks with new applications. 
This work would contribute to advancing the community's possibilities, reducing the workload for researchers to exploit DL,  allowing Neuroscience to benefit of its most recent advancements.

% Fabrice
% Outline of the rest paper ???
% The rest of the paper is organised as hereby outlined:
% first a historical perspective of the rise of Deep Learning, then,
% a general presentation of the vast field of Neuroscience, followed 
% by a definition of Neuroinformatics and the role open-source culture
% and Deep Learning do and would play in it;
% subsequently the methodology to collect and present the libraries collection is described,
% and the most prominent features are discussed;
% lastly, discussions and final remarks are offered to the readers.

%\newpage
\section{Background}
 \subsection{Deep Learning}
Deep Learning (DL) has contributed many best solutions to problems in its parent field, Machine Learning, thanks to theoretical and technological achievements that unlocked its intrinsic versatility.\\
Machine Learning is the study of computer algorithms that tackle problems without complete access to predefined rules or analytical, closed-form solutions. \\
The algorithms often require a training phase to adjust parameters and satisfy internal or external constraints (e.g. of exactness, approximation or generality) on dedicated data for which solutions might be already known.\\
Machine Learning comprises a wide array of statistical and mathematical methods, including Artificial Neural Networks (ANNs), biologically inspired systems that connect inputs and outputs through simple computing units (neurons), which act as function approximators.\\ 
Each unit implements a nonlinear function of the weighted sum of its inputs, thus the output of the whole ANN is a composite function, as formally intended in mathematics.
The networks of neurons are most often layered and "feed-forward", meaning that units from any layer only output results to units in subsequent layers.
The width of a layer refers to its neuron count, while the depth of a network refers to its layer count.
The typical architecture instantiating the above characteristics is the MultiLayer Perceptron~\cite{rosenblatt1958perceptron} (MLP).\\
Universal approximation theorems~\cite{cybenko_approximation_1989}~\cite{hornik_multilayer_1989} %XXXREF 
ensure that, whenever a nonlinear network as the MLP is either bound in width and unbound in depth or viceversa, its weights can then be set to represent virtually any function (i.e. a wide variety of functions families). \\
The training problem thus consists in building networks with sets of weights so to instantiate or approximate the function that would solve the assigned task, or that represents the input-output relation.
This search is not trivial: it can be framed as the optimization problem for a functional over the ANN weights. 
Such functional, typically called "loss function", associates the "errors" made on the training data to the neural net parameters (its weights), acting as a total performance score.
Approaching local minima of the loss function and improving the network performance on the training data is the prerequisite to generalize on real world and unseen data.\\
DL is concerned with the use of deep ANNs, namely characterized by depth, stacking several intermediate, (hidden) layers between input and output units. \\
As mentioned above, other dimensions being equal, depth increases the representational power of ANNs and, more specifically, aims at modeling complicated functions as meaningful compositions of simpler ones. \\
As with their biological counterparts~\cite{fukushima_neocognitron_1980}, depth is supposed to manage hierarchies of features from larger input portions, capturing characteristics often inherent to real world objects and effective in modeling actual data.\\
Overall, depth is one of the key features that allowed to overcome historical limits~\cite{minsky69perceptrons} of simpler ANNs such as the Perceptron. 
At the same time, depth comes with numerical and methodological hardships in models training. \\
Part of the difficulties arise as the search space for the optimal set of parameters grows considerably with the number of layers (and their width as well). \\
Other issues are strictly numerical, since the training algorithms include long computation chains that may affect the stability of training and learning.\\ 
Hence, new or rediscovered ideas in training protocols and mathematical optimization (e.g. applying the "backpropagation of errors" algorithm to neural nets~\cite{Rumelhart:1986we})
played an important role through times when the scientific interest and hopes in ANNs faded (so called "AI winters"), paving the way for later advancement.\\
The main drivers for the latest success of deep neural networks are of varied nature, and can be schematised as technical and human related factors.\\
On a technical side DL has profited from~\cite{meijering_birds-eye_2020}: 
\begin{itemize}
    \item the datafication of the world, i.e. the growing availability of (Big) data
    \item the diffusion of Graphical Processing Units (GPUs) as hardware tools.
\end{itemize}
To outperform classic machine learning models, deep neural networks often require larger quantities of data samples.
Such data hunger and high parameters count contribute to the high requirements of deep models in terms of memory, number of operations and computation time.
Training models with highly parallelized and smartly scheduled computations gained momentum thanks to GPUs.\\
In 2012 a milestone exemplified both the above technical aspects, when AlexNet~\cite{NIPS2012_4824}, a deep Convolutional Neural Network (CNN) based on ideas from Fukushima~\cite{fukushima_neocognitron_1980} and LeCun~\cite{LeCunBoserDenkerEtAl89} -~\cite{LeCun98gradient-basedlearning}, won the ImageNet Large Scale Visual Recognition Challenge after being trained using two GPUs~\cite{russakovsky2014imagenet}.
Since then, Deep Learning has brought new outstanding results in various tasks and domains, processing different data types. 
Deep networks can nowadays work on image, video, audio, text, and speech data, time series and sequences, graphs, and more;
the main tasks consist in classification, prediction, or estimating the probability density of data distributions, with the possibility of modifying, completing the input or even generating new instances. \\
On a more sociological side, the drivers of Deep Learning success can be related to the synergy of big tech companies, advanced research centers, and developer communities~\cite{valliani_deep_2019}. 
Investments of economical and scientific resources in relatively independent, collective projects, such as open-source libraries, frameworks, and APIs (Application Programming Interfaces), have offered varied tools adapted to multiple specific situations and objectives, exploiting horizontal organization~\cite{raymond2001cathedral} and mixing top-down and bottom-up approaches. 
%Mnf:~\cite{noauthor_cathedral_nodate} è corretto? Se è un riferimento incerto o poco difendibile, meglio rimuoverlo.
%Re: il riferimento è corretto anche se "noauthor, nodate" sembrano tag incerti, in realtà sono presenti: è un saggio influente 
%sul successo dello sviluppo dei sistemi Linux e open, considerato anche un manifesto
%Si affianca al riferimento di Valliani, posso mettere due citazioni testuali e avvicinarle
It is difficult to imagine a rapid rise of successful endeavors, without both active communities and the technical means to incorporate and manage lower-level aspects. \\
In fact, applying Deep Learning to a relevant problem in any research field requires, in addition to specific domain knowledge, a vast background of statistical, mathematical, and programming notions and skills. 
The tools that support scientists and engineers in focusing on their main tasks encompass the languages to express numerical operations on GPUs, such as CUDA~\cite{cuda} and cuDNN~\cite{Chetlur2014cuDNNEP} by NVIDIA, as well as the frameworks to design models, like TensorFlow~\cite{tensorflow2015-whitepaper} and Keras~\cite{chollet2015keras} by Google, and PyTorch by Meta~\cite{NEURIPS2019_9015}, or the supporting strategies to build data pipelines.\\
Many Deep Learning achievements are relevant to biomedical and clinical research, and the above presented tools have enabled explorations of the capabilities of deep neural networks with neuroscience and biomedical data. \\
A fuller exploitation and routinely employment of modern algorithms are yet to come, both in research and clinical practice. 
This process would accelerate by popularizing, democratizing, and jointly developing models, improving their usability, and expanding their environments, i.e. by wrapping solutions into libraries and shared frameworks.

\subsection{Neuroscience}
As per the Nature journal, <<Neuroscience is a multidisciplinary science that is concerned with the study of the structure and function of the nervous system. It encompasses the evolution, development, cellular and molecular biology, physiology, anatomy and pharmacology of the nervous system, as well as computational, behavioural and cognitive neuroscience>>~\cite{nature}.\\
Expanding, neuroscience investigates:
\begin{itemize}
    \item the evolutionary and individual development of the nervous system;
    \item the cellular and molecular biology that characterizes neurons and glial cells;
    \item the physiology of living organisms and the role of the nervous system in the homeostatic function;
    \item the anatomy, i.e. the identification and description of the system's structures;
    \item  pharmacology, i.e. the effect of chemicals of external origin on the nervous system, their interactions with endogenous molecules;
    \item the computational features of the brain and nerves, how information is processed, which mathematical and physical models best predict and approximate the behaviour of neurons;
    \item cognition, the mental processes at the intersection of psychology and computational neuroscience;
    \item behaviour as a phenomenon rooted in genetics, development, mental states, and so forth.
\end{itemize}
The techniques to access tissues and structures of the nervous system are often shared by disciplines focused on other physiological systems, and some of these processes have been computer aided for long.\\
Moreover, nerve cells have distinctive electromagnetic properties and their activity directly and indirectly generates detectable signals, adding physical and technical specificity to Neuroscience. \\
Overall, neuroscience research is profoundly multi-modal. 
Data are managed and processed inside a model depending on their type and format. The most prominent categories of data involved in neuroscience research comprise 2,3-D images or video on the one side, and sequences or signals on the other. Still it is important to acknowledge the different phenomena, autonomous or provoked by the measurement apparatus, underlying data generation and acquisition. 
Bioimages may be produced from:
\begin{itemize}
    \item Magnetic Resonance Imaging (MRI)
    \item X-rays
    \item Tomography with different penetrating waves
    \item Histopathology microscopy
    \item Fundus photography (retinal images)
\end{itemize}
and more.\\
Neuroscience sequences may come from:
\begin{itemize}
    \item Electromiography (EMG)
    \item Electroencephalography (EEG)
    \item Natural language, text records
    \item Genetic sequencing
    \item Eye-tracking
\end{itemize}
and more.\\
Adding to the above, other data types are common in neuroscience, e.g. tabular data, text that may come from medical records written by physicians for diagnostic purposes, test scores, inspections of cognitive and sensorimotor functions, as the National Institute of Health (NIH) Stroke Scale test scores~\cite{nihss}, and more broadly clinical reports from anamneses or surveys.

%Mnf 
\subsection{Neuroinformatics}
%%%%%%%%%%%%%%%%%%%%%%%%%%%%%%%%%
% NEUROINFORMATICS: DATA AND SOFTWARE
Neuroscience is evolving into a data-centric discipline.
Modern research heavily depends on human researchers as well as machine agents to store, manage and process computerized data from the experimental apparatus to the end stage.\\
Before delving in the specifics of artificial neural networks applied to the study of biological neural systems, it is useful to outline the broader concepts of Neuroinformatics, regarding data and coding, especially in the light of open culture.\\
According to the International Neuroinformatics Coordinating Facility (INCF), <<Neuroinformatics is a research field devoted to the development of neuroscience data and knowledge bases together with computational models and analytical tools for sharing, integration, and analysis of experimental data and advancement of theories about the nervous system function. In the INCF context, neuroinformatics refers to scientific information about primary experimental data, ontology, metadata, analytical tools, and computational models of the nervous system. The primary data includes experiments and experimental conditions concerning the genomic, molecular, structural, cellular, networks, systems and behavioural level, in all species and preparations in both the normal and disordered states>>~\cite{incf1}.
% FAIR DATA AND INCF
Given the relevance of Neuroinformatics to Neuroscience, supporting open and reproducible science implies and requires attention to standards and best practices regarding open data and code. \\
The INCF itself is an independent organization devoted to validate and promote such standards and practices, interacting with the research communities~\cite{abrams_standards_2021} and aiming at the "FAIR principles for scientific data management and stewardship"~\cite{wilkinson_fair_2016}. \\
FAIR principles consist in:
\begin{itemize}
    \item being Findable, registered and indexed, searchable, richly described in metadata;
    \item being Accessible, through open, free, universally implementable protocols;
    \item being Interoperable, with appropriate standards for metadata in the context of knowledge representation;
    \item being Reusable, clearly licensed, well described, relevant to a domain and meeting community standards.
\end{itemize}
% FAMOUS SOURCES
%Mnf: Io inserirei un paragrafo tipo quella sotto, in cui menzionerei alcuni tra i più noti software in ambito di neuroscienze, e.g. SPM, Freesurfer, FSL, + altri che non conosco in ambito di EEG. Per l'EMG si può includere mettere  PAWFE volendo, anche se è limitatamente famoso). È da inserire per ognuno un riferimento e un link.
%Fabrice direi Freesurfer, FSL, 3D Slicer, SPM, EEGLAB, Brainstorm, MNE, PAWFE 
Among free and open resources, several software and organized packages integrating pre-processing and data analysis workflows for neuroimaging and signal processing became the reference for worldwide researchers in Neuroscience. \\
Such tools allow to perform scientific research in neuroscience easily in solid and repeatable ways. 
It can be useful to mention, for neuroimaging, Freesurfer\footnote{\url{https://surfer.nmr.mgh.harvard.edu/}}~\cite{fischl2012freesurfer} and FSL\footnote{\url{https://fsl.fmrib.ox.ac.uk/fsl/fslwiki}}~\cite{Smith2004AdvancesIF} that are standalone softwares, and the MATLAB-connected SPM\footnote{\url{https://www.fil.ion.ucl.ac.uk/spm/}}~\cite{spm_sholarpedia}. In the domain of signal processing, examples are EEGLAB\footnote{\url{https://sccn.ucsd.edu/eeglab/index.php}}~\cite{delorme_eeglab_2004}, Brainstorm\footnote{\url{https://neuroimage.usc.edu/brainstorm/Introduction}}~\cite{tadel_brainstorm_2011}, PaWFE\footnote{\url{http://ninapro.hevs.ch/node/229}}~\cite{atzori_pawfe_2019}, all MATLAB related yet free and open, and MNE\footnote{\url{https://mne.tools/stable/index.html}}~\cite{GramfortEtAl2013a}, that runs on Python. Regarding applications for neurorobotics and Brain Computer Interfaces (BCIs), a recent opensource platform can be found in ROS-neuro\footnote{\url{https://github.com/rosneuro}}~\cite{rosneuro}.\\
% OPEN SOFTWARE FOR ANALYSIS IN NEUROSCIENCE
The interested readers can find lists of open resources for computational neuroscience (including code, data, models, repositories, textbooks, analysis, simulation and management software) at Open Computational Neuroscience Resource \footnote{\url{https://github.com/asoplata/open-computational-neuroscience-resources}} (by Austin Soplata), and at Open Neuroscience \footnote{\url{https://open-neuroscience.com/}}.
Additional software resources oriented to Neuroinformatics in general, but not necessarily open, can also be found as indexed at "COMPUTATIONAL NEUROSCIENCE on the Web" \footnote{\url{https://compneuroweb.com/sftwr.html}} (by Jim Perlewitz).

\subsection{Bringing Deep Learning to the Neurosciences}
%%%%%%%%%%%%SCHEMA NECESSARIO
% NEUROSC AND DL HAVE FRAIL THEORY
% DL AND CODING ARE HARD AND DEEP TOPICS
% DL IS ALREADY QUITE OPEN FRIENDLY
% THERE IS A NEED FOR EASY AND OPEN NEURO-DL

% DL AND OPEN SCIENCE
% DL AND STANDARDS
% DL AND NEUROSC

The Deep Learning community is accustomed to open science, as many datasets, models, programming frameworks and scientific outcomes are  publicly released by both academia and companies continuously.
However, while Deep Learning can openly provide state-of-the-art models to old and new problems in Neuroscience, theoretical understanding, formalization and standardisation are often yet to be achieved, which may prevent adoption in other research endeavors.
From a technical standpoint, deep networks are a viable tool for many tasks involving data from the brain sciences. 
Image classification has arguably been the task in which deep neural networks have had the highest momentum, in terms of pushing the state of the art forward. This translates now in a rich taxonomy of architectures and pre-trained models that consistently maintain interesting performances in pattern recognition, across a number of image domains.\\
Pattern recognition is indeed central for diagnostic purposes, in the form of classification of images with pathological features (e.g. types of brain tumors or meningiomas), segmentation of structures (such as the brain, brain tumors or stroke lesions), classification of signals (e.g. classification of electromyography or electro encephalography data), as well as for action recognition in Human-Computer Interfaces (HCIs). The initiatives BRain Tumor Segmentation (BRATS) Challenge\footnote{\url{https://www.med.upenn.edu/cbica/brats/}}~\cite{brats}, Ischemic Stroke LEsion Segmentation (ISLES) Challenge\footnote{\url{https://www.isles-challenge.org/}}~\cite{isles}-~\cite{isles2}, and Ninapro\footnote{\url{http://ninaweb.hevs.ch/node/7}}~\cite{Atzori2012BuildingTN} are examples of data releases for which above-mentioned tools proved effective.\\
There are models learning image-to-image functions, capable of enhancing data, preprocessing it, correcting artifacts and aberrations, allowing smart compression as well as super-resolution, and even expressing cross-modal transformations between different acquisition apparatus.\\
In the related tasks of object tracking, action recognition and pose estimation, research results from the automotive sector or crowd analysis have inspired solutions for behavioural neuroscience, especially in animal behavioral studies.\\
When dealing with sequences, deep networks success in Computer Vision has inspired CNN-based approaches to EEG and EMG studies \cite{park_movement_2016} - \cite{atzori_deep_2016}, either with or without relying on 2D data, given that mathematical convolution has a 1D version, and 1D signals have 2D spectra.
Other architectures more directly instantiate temporal and sequential aspects, e.g. Recurrent Neural Networks (RNNs) such as the Long Short Term Memory (LSTM)~\cite{HochSchm97} and Gated Recurrent Units (GRUs) \cite{cho-etal-2014-properties}, and they too can be applied to sequence problems and sub-tasks in neuroscience, such as decoding time-dependent brain signals.\\
Although deep neural network do not explicitly model the nervous system, they are inspired by biological knowledge and mimic some aspects of biological computation and dynamical systems.
This has inspired new comparative studies, and analogy approaches to learning and perception, in a unique way among machine learning algorithms~\cite{yamins_using_2016}.\\%\cite{rnnsurvey}
Many neuroinformatic studies demonstrate how novel deep learning concepts and methods apply to neurological data~\cite{valliani_deep_2019}.
However, they often showcase new further achievements in performance metrics that do not translate directly to new accepted neuroscience discoveries or clinical best practices.\\
Such results are very often published together with open code repositories, allowing reproducibility, yet they may not be explicitly organized for widespread routinely adoption in domains different from machine learning. 
Algorithms are usually written in open programming languages like Python~\cite{python}, R~\cite{Rlang}, Julia~\cite{Julia-2017}, and deep learning design frameworks such as TensorFlow, PyTorch or Flux~\cite{innes:2018}. 
Still, they are more inspiring to the experienced machine learning researcher, rather than practically helpful to end-users such as neuroscientists.\\
In fact, to successfully build a deep learning application from scratch, a vast knowledge is needed in the data science aspect of the task and in coding , as much as in the theoretical and experimental foundations and frontiers of the application domain, here being Neuroscience.\\
For the above reasons, the open source and open science domains are promising frames for common development and testing of relevant solutions for Neuroscience, as they provide an active flow of ideas and robust diversification, avoiding "reinvention of the wheel", harmful redundancies or starting from completely blank states. \\
As a contribution in clarifying the current situation and reducing the workload for researchers, this work collects and analyzes several open libraries that implement and facilitate Deep Learning application in Neuroscience, with the aim of allowing worldwide scientists to identify the most suitable options for their inquiries and clinical tasks.

\section{Methods}
The large corpus of available open code makes useful to specify what qualifies as a coding library or a framework, rather than as a model accompanied by utilities, for the present scope.\\ 
In programming, a library is a collection of pre-coded functions and object definitions, often relying on one another, and written to optimize programming for custom tasks.
%These tasks require behaviors that one can implement calling functions from the library. 
The functions are considered useful and unmodified across multiple unrelated programs and tasks. 
The main program at hand calls the library, in the control flow specified by the end-users.\\
A framework is a higher level concept, akin to the library, but typically with a pre-designed control flows in which custom code from the end-users is inserted.\\
For instance, a repository that simply collects the functions that define and instantiate a deep model would not be considered a library. 
On the other hand, collections of notebooks that allow to train, retrain and test models with several architectures, while possibly taking care also of data pre-processing and preparation, would be considered libraries (and frameworks) for the present scopes.
The explicit definition of the authors, their aims and their maintainance of the library is relevant as well, in determining if a repository would be considered a library, toolkit, toolbox, or other.\\

For the sake of the review, several resources were queried or scanned. 
Google Scholar was queried with:
\begin{itemize}
    \item "deep learning library" OR "deep learning toolbox" OR "deep learning package" -"MATLAB deep learning toolbox" -"deep learning toolbox MATLAB"
%Mnf: non mi è chiaro quali sono le parole cercate in scholar
%Mnf: magari farei una ricerca aggiuntiva con "deep neural networks xxx" (dove al posto di xxx si mette toolbox, package, etc, sia per Scholar, che per Pubmed e i portali. (mi aspetterei che i risultati non cambino molto e renderebbe più solida la parte dei metodi.
\end{itemize}
preserving the top 100 search results, ordered for relevance by the engine algorithm.
On PubMed the queries were:
\begin{itemize}
    \item opensource (deep learning) AND (toolbox OR toolkit OR library);
    \item (EEG OR EMG OR MRI OR (brain (X-ray OR CT OR PT))) (deep learning) AND (toolbox OR toolkit OR library).
\end{itemize}
 Moreover, the site \url{https://open-neuroscience.com/} was scanned specifically for "deep learning" mentions, and relevant papers cited or automatically suggested throughout the query process were considered for evaluation, as well as the platform of the Journal of Open Source Software at \url{https://joss.theoj.org/}.

The collected libraries were organized according to the principal aim, in the form of data type processed, or the supporting function in the workflow, thus dividing:
\begin{enumerate}
    \item libraries for sequence data (e.g. EMG, EEG)
    \item libraries for image data (including scalar volumes, 4-dimensional data as in fMRI, video)
    \item libraries and frameworks to support model building, evaluation, data ingestion
\end{enumerate}
In each category, a set of three tables present separately the results related to the following libraries characteristics:
\begin{enumerate}
    \item domain of application
    \item model engineering
    \item technology and sources
\end{enumerate}
The domain of application comprises the \textbf{Neuroscience area}, the \textbf{Data types} handled, the provision of \textbf{Datasets}, and the machine learning \textbf{Task} to which the library is dedicated.\\
The model engineering tables include informations on the architecture of DL \textbf{Models} manageable in the library, the \textbf{DL framework} and \textbf{Programming language} main dependencies, and the possibility of \textbf{Customization} for the model structure or training parameters.\\
Technology and sources refer to the type of \textbf{Interface} available for a library, whether it works \textbf{Online//Offline}, specifically with real-time data or logged data. \textbf{Maintenance} refers to the ongoing activity of releasing features, solving issues and bugs or offering support through channels, \textbf{Source} specifies where code files and instructions are made available.

\section{Results: Deep Learning Libraries}
%Mnf: Sarebbe importante inserire un paragrafo riassuntivo, che sintetizzi tutti i risultati in poche righe (e.g. 5-6). Ad esempio: 
%The analysis of the literature allowed to identify a total of XXX libraries aiming at the application of deep learning to neuroscience.
%Among those, XXX libraries were discarded from the analysis because ... .

The analysis of the literature allowed to select a total of 48 publications describing libraries that implement or empower deep learning applications for neuroscience.
Despite open source and effectiveness, several publications did not provide an ecosystem of reusable functions. Proofs of concept and single-shot experiments were discarded.

\subsection{Libraries for sequence data}
Libraries and frameworks for sequence data are shown in Tables~\ref{tab:tabseq1} (domains of application),~\ref{tab:tabseq2} (models characteristics),~\ref{tab:tabseq3} (technologies and sources).
The majority of process EEG sygnals, which are among the most common types of sequential data in Neuroscience research.
A common objective is deducing the activity or state of the subject, based on temporal or spectral (2D) patterns. 
Deep Learning is capable of bypassing some of the preprocessing steps often required by other common statistical and engineering techniques, and it comprises both 1D and 2D approaches, through MLPs, CNNs or RNNs architectures. 
\textsf{BioPyC} is an example of such scenario. It offers the possibility to train a pre-set CNN architecture as well as loading and training a custom model. Moreover, It can process different types of sequence data, making it very versatile and applicable/ suitable/usable in/for different neuroscience area.
Another example of sequence-oriented library is \textsf{gumpy}, whose intended area of application is that of Brain Computer Interfaces (BCIs), where decoding a signal is the first step towards communication and interaction with a computer or robotic system. Given the setting, \textsf{gumpy} allows working with EEG or EMG data and suits them with specific defaults, e.g. 1-D CNNs, or LSTMs.\\
Notable mentions in the sequence category are the library \textsf{Traja}  and the \textsf{VARDNN toolbox}, as they depart from the common scenarios of previous examples.
\textsf{Traja} stands out as an example of less usual sequential data, namely trajectory data (sequences of coordinates in 2 or 3 dimensions, through time). Moreover, in \textsf{Traja} sequences are modeled and analyzed employing the advanced architectures of Variational AutoEncoders (VAEs) and Generative Adversarial Networks (GANs), usually encountered in image tasks. With different theoretical backgrounds, both architectures allow simulation and characterization of data through their statistical properties. 
The \textsf{VARDNN toolbox} allows analyses on BOLD signals, in the established domain of functional Magnetic Resonance Imaging (fMRI), but uses a unique approach to autoregressive processes mixed with deep neural networks, allowing to perform causal analysis and to study functional connections between brain regions through their patterns of activity in time.\\
%Overall, the libraries oriented to sequence data analysis are mainly directed at classification of EEG signals, which have a variety of acquisition settings and downstream applications, that could be largely approached with the aid of deep models as a part of the pipeline.

\newpage
\clearpage
\onecolumn
\definecolor{lavender}{rgb}{0.9, 0.9, 0.98}

% &&&&&&&&&&&&&&

\begin{landscape}
%%\begin{table}[h]
        \centering
        
        \small \sffamily
        \begin{tabularx}{600pt}{*{5}{lX} }
            \toprule
            \textbf{Name} & \textbf{Neuroscience area} & \textbf{Data type} & \textbf{Datasets} & \textbf{Task}\\
            
            \midrule
            \rowcolor{lavender}
            BioPyC \cite{appriou_biopyc_2021} & General & Sequences (EEG, miscellaneous) & No & Classification \\
braindecode \cite{schirrmeister_deep_2017} & General & Sequences (EEG, MEG) & External & Classification \\\rowcolor{lavender}
DeLINEATE \cite{kuntzelman_deep-learning-based_2021} & General & Images, sequences & External & Classification \\
EEG-DL \cite{hou_novel_2020} & BCI & Sequences (EEG) & No & Classification \\\rowcolor{lavender}
gumpy \cite{tayeb_gumpy_2018-1} & BCI & Sequences (EEG, EMG) & No & Classification \\
DeepEEG & Electrophysiology & Sequences (EEG) & No & Classification \\\rowcolor{lavender}
ExBrainable \cite{huang_exbrainable_2022} & Electrophysiology & Sequences (EEG) & External & Classification, XAI \\
Traja \cite{shenk_traja_2021} & Behavioural neuroscience & Sequences (Trajectory coordinates over time) & No & Prediction, Classification, Synthesis \\\rowcolor{lavender}
VARDNN toolbox \cite{okuno_vector_2021} toolbox & Connectomics (Functional Connectivity) & Sequences (BOLD signal) & No & Time series  causal analysis \\
\\
            
            \bottomrule
        
        \end{tabularx}
        \captionof{table}{Domains of applications for the libraries and frameworks processing sequence data}%\rmfamily
        \label{tab:tabseq1}
    %\end{table}
\end{landscape}

\begin{landscape}
%\begin{table}[h]
        \centering
        \small \sffamily
        \begin{tabularx}{600pt}{*{5}{lX} }
            \toprule
            \textbf{Name} & \textbf{Models} & \textbf{DL framework} & \textbf{Customization} & \textbf{Programming language}\\
            \midrule
            \rowcolor{lavender}
            BioPyC  & 1-D CNN & Lasagne & Yes (weights, model) & Python\\
braindecode  & 1-D CNN & PyTorch & Yes (weights, model) & Python\\ \rowcolor{lavender}
DeLINEATE  & CNN  & Keras, TensorFlow & Yes (weights, model) & Python\\
EEG-DL & Miscellaneous & TensorFlow & Yes (weights, model) & Python, MATLAB\\ \rowcolor{lavender}
gumpy & CNN, LSTM & Keras, Theano & Yes (weights, model) & Python\\
DeepEEG & MLP, 1,2,3-D CNN, LSTM & Keras, TensorFlow & Yes (weights) & Python\\ \rowcolor{lavender}
ExBrainable & CNN & PyTorch & Yes (weights) & Python\\
Traja & LSTM, VAE, GAN & PyTorch & Yes (weights, model) & Python\\ \rowcolor{lavender}
VARDNN toolbox & Vector Auto-Regressive DNN  & Deep Learning Toolbox (MATLAB) & Yes (weights) & MATLAB\\
            
            \bottomrule
        \end{tabularx}
        \captionof{table}{Model engineering specifications for the libraries and frameworks processing sequence data}
        %\rmfamily
        \label{tab:tabseq2}
%\end{table}

\end{landscape}

\begin{landscape}
%\begin{table}[h]
        \centering
        \small \sffamily
        \begin{tabularx}{600pt}{*{5}{lX} }
            \toprule
            \textbf{Name} & \textbf{Interface} & \textbf{Online/Offline} & \textbf{Maintenance} & \textbf{Source}\\
            \midrule
            \rowcolor{lavender}
            BioPyC & Jupyter Notebooks & Offline & Active & \url{gitlab.inria.fr/biopyc/BioPyC/} \\
braindecode & None & Offline & Active & \url{github.com/braindecode/braindecode} \\ \rowcolor{lavender}
DeLINEATE  & GUI, Colab Notebooks & Offline & Active & \url{bitbucket.org/delineate/delineate} \\
EEG-DL & None & Offline & Active & \url{github.com/SuperBruceJia/EEG-DL} \\ \rowcolor{lavender}
gumpy & None & Online, Offline & Inactive & \url{github.com/gumpy-bci} \\
DeepEEG & Colab Notebooks & Offline & Inactive & \url{github.com/kylemath/DeepEEG} \\ \rowcolor{lavender}
ExBrainable & GUI  & Offline & Active & \url{github.com/CECNL/ ExBrainable} \\
Traja & None & Offline & Active & \url{github.com/traja-team/traja} \\ \rowcolor{lavender}
VARDNN toolbox & None & Offline & Active & \url{github.com/takuto-okuno-riken/vardnn} \\
            \\
            
            \bottomrule
        \end{tabularx}
        \captionof{table}{Technological aspects and code sources for the libraries and frameworks processing sequence data}
        %\rmfamily
        \label{tab:tabseq3}
%\end{table}
\end{landscape}

\clearpage
%\twocolumn

\subsection{Libraries for image data}
Libraries and frameworks for image data are shown in Tables ~\ref{tab:tabimg1} (domains of application),~\ref{tab:tabimg2} (models characteristics),\ref{tab:tabimg3} (technologies and sources).
Computer vision and 2D image processing are arguably the fields in which DL has achieved the most impressive and state-of-art defining results, often inspiring and translating breakthroughs in other domanis.
Classification and segmentation (i.e. the separation of parts of the image based on their classes) are the most common tasks addressed by the image processing libraries. 
Magnetic resonance is the primary source of data; however, various deep learning libraries are built microscopic and eye-tracking data as well.
Most of the libraries collected in our analysis take advantage of classical CNN architectures for classification, Convolutional AutoEncoders (CAEs) for segmentation, and GANs for synthesis.
It is common to employ transfer learning to lessen the computational and memory burden during the training phase, and take advantage of pre-trained models. Transfer learning consists in initializing models with parameters learnt on usually larger data sets, possibly from different domains and tasks, with varying amounts of further training in the target domain.
The best such examples are pose-estimation libraries extending the \textsf{DeepLabCut} system, arguably the most relevant project on the topic.
\textsf{DeepLabCut} is an interactive framework for labelling, training, testing and refining models, that originally exploits the weights learned from ResNets (or newer architectures) on the ImageNet data. 
The results match human annotation using quite few training samples, holding for many (human and non-human) animals, and settings. 
The documentation and demonstrative notebooks and tools offered by the Mathis Lab allow different levels of understanding and customization of the process, with high levels of robustness.
Among the considered libraries, two set apart from the majority given the type of tasks they perform: \textsf{GaNDLF} addresses eXplainable AI (XAI), i.e. Artificial Intelligence whose decisions and outputs can be understood by humans through more transparent mental models; \textsf{ANTsX} performs both the co-registration step and super-resolution as a quality enhancing step for neuroimages, with the former being usually performed by traditional algorithms. 
\textsf{GaNDLF} sets its goal as the provision of deep learning resources in different layers of abstraction, allowing medical researchers with virtually no ML knowledge to perform robust experiments with models trained on carefully split data, with augmentations and preprocessing, under standardized protocols that can easily integrate interpretability tools such as Grad-CAM~\cite{gradcam}
and attention maps, which highlight the parts of an image according to how they influenced a model outcome.
The \textsf{ANTsX} ecosystem is of similar wide scope, and is intended to build workflows on quantitative biology and medical imaging data, both in Python and R languages. Packages from the same ecosystem perform registration of brain structures (by classical methods) as well as brain extraction by deep networks.
\newpage
\clearpage
\onecolumn
\definecolor{lavender}{rgb}{0.9, 0.9, 0.98}

% &&&&&&&&&&&&&&

\begin{landscape}
%\begin{table}
        \centering
        
        \small \sffamily
        \begin{tabularx}{600pt}{*{5}{lX} }
            \toprule
            \textbf{Name} & \textbf{Neuroscience area} & \textbf{Data type} & \textbf{Datasets} & \textbf{Task}\\
            
            \midrule
            \rowcolor{lavender}
            AxonDeepSeg \cite{zaimi_axondeepseg_2018} & Microbiology, Histology & Img (SEM, TEM) & External & Segm. \\ 
            DeepCINAC \cite{denis_deepcinac_2020} & Electrophys. & Vid (2-photon calcium) & No & Class. \\ \rowcolor{lavender}
DeepLabCut \cite{nath_using_2019} & Behavioral  neuroscience & Vid & No & Pose est. \\
DeepNeuro \cite{beers_deepneuro_2021} & Neuroimaging & Img  (fMRI, misc.) & No & Class.,  Segm.,   Synthesis \\ \rowcolor{lavender}
DeepVOG \cite{yiu_deepvog_2019} & Oculography & Img, Vid & Demo & Segm. \\ 
DeLINEATE \cite{kuntzelman_deep-learning-based_2021} & General & Img,  sequences & External & Class. \\ \rowcolor{lavender}
DNNBrain \cite{chen_dnnbrain_2020} & Brain mapping & Img & No & Class. \\ 
ivadomed \cite{gros_ivadomed_2020} & Neuroimaging & Img (2D, 3D) & No & Class.,  Segm. \\ \rowcolor{lavender}
MEYE \cite{mazziotti_meye_2021} & Oculography & Img, Vid & Yes & Segm. \\ 
Allen Cell  Structure Segmenter \cite{chen_allen_2018} & Microbiology,  Histology & Img  (3D-fluor. microscopy) & No & Segm. \\ \rowcolor{lavender}
VesicleSeg \cite{imbrosci_automated_2022} & Microbiology,  Histology & Img (EM) & No & Segm. \\ 
CDeep3M2 \cite{haberl_cdeep3mplug-and-play_2018} & Microbiology,  Histology & Img  (misc. microscopy) & Yes & Segm. \\ \rowcolor{lavender}
CASCADE \cite{rupprecht_database_2021} & Electrophys. & Vid  (2-photon calcium),  Seq & Yes & Event detection \\ 
ScLimibic \cite{greve_deep_2021} & Neuroimaging & Img (MRI) & External & Segm. \\ \rowcolor{lavender}
ALMA \cite{aljovic_deep_2022} & Behavioral  neuroscience & Vid & External & Pose est.,  Class. \\ 
fetal-code \cite{rutherford_automated_2021} & Neuroimaging & Img  (rs-fMRI) & External & Segm. \\ \rowcolor{lavender}
ClinicaDL \cite{thibeau-sutre_clinicadl_2022} & Neuroimaging & Img (MRI, PET) & External & Class.,  Segm. \\ 
DeepNeuron \cite{zhou_deepneuron_2018} & Microbiology, Histology & Img  (confocal microscopy) & No & Obj. detect.,  Segm. \\ \rowcolor{lavender}
GaNDLF \cite{pati_gandlf_2021} & Medical Imaging & Img (2D, 3D) & External  & Segm., Regression, XAI \\ 
MesoNet \cite{xiao_mesonet_2021} & Neuroimaging & Img (fluoresc. microscopy) & External  & Segm., Registration \\ \rowcolor{lavender}
MARS, BENTO \cite{segalin_mouse_2021} & Behavioral neuroscience & Vid & Yes & Pose est., Class., Action rec., Tag \\ 
NiftyNet \cite{gibson_niftynet_2018} & Medical Imaging & Img (MRI, CT) & No & Class., Segm., Synth. \\ 
ANTsX \cite{tustison_antsx_2021} (ANTsPyNet, ANTsRNet) & Neuroimaging & Img (MRI) & No & Classificastion, Segm., Registr., Super-res. \\ \rowcolor{lavender}
MARS, BENTO \cite{segalin_mouse_2021} & Behavioral neuroscience & Vid & Yes & Pose est., Class., Action rec., Tag \\ 
Visual Fields Analysis \cite{josserand_visual_2021} & Eye tracking, Behavioral neuroscience & Vid & No & Pose est., Class.    \\          
            \bottomrule
        \end{tabularx}
        \captionof{table}{Domains of applications for the libraries and frameworks processing image data}
        \label{tab:tabimg1}
    %\end{table}
\end{landscape}

\begin{landscape}
%\begin{table}[h]
        \centering
        \small \sffamily
        \begin{tabularx}{600pt}{*{5}{lX} }
            \toprule
            \textbf{Name} & \textbf{Models} & \textbf{DL framework} & \textbf{Customization} & \textbf{Programming language}\\
            \midrule
             \rowcolor{lavender}
             AxonDeepSeg & CAE & TensorFlow & Yes (weights) & Python \\ 
DeepCINAC & DeepCINAC (CNN+LSTM) & Keras, TensorFlow & Yes (weights) & Python \\  \rowcolor{lavender}
DeepLabCut & CNN & TensorFlow & Yes (weights) & Python \\ 
DeepNeuro & CNN, CAE, GAN & Keras, TensorFlow & Yes (weights, model) & Python \\  \rowcolor{lavender}
DeepVOG & CAE & TensorFlow & No & Python \\ 
DeLINEATE & CNN  & Keras, TensorFlow & Yes (weights, model) & Python \\  \rowcolor{lavender}
DNNBrain & CNN & PyTorch & Yes (model) & Python \\ 
ivadomed & 2,3-D CNN, CAE & PyTorch & Yes (weights, model) & Python \\  \rowcolor{lavender}
MEYE & CAE, CNN & TensorFlow & Yes (model) & Python \\ 
Allen Cell Structure Segmenter & CAE & PyTorch & No & Python \\  \rowcolor{lavender}
VesicleSeg & CNN & PyTorch & No & Python \\ 
CDeep3M2 & CAE & TensorFlow & Yes (weights) & Python \\  \rowcolor{lavender}
CASCADE & 1-D CNN & TensorFlow & Yes (weights) & Python \\ 
ScLimibic & 3-D CAE & neurite, TensorFlow & No & Python \\  \rowcolor{lavender}
ALMA & CNN & Unspecified & No & Python \\ 
fetal-code & 2-D CNN & TensorFlow & No & Python \\  \rowcolor{lavender}
ClinicaDL & CNN, CAE  & PyTorch & Yes & Python \\ 
DeepNeuron & CNN & Unspecified & No & C++ \\  \rowcolor{lavender}
GaNDLF & CNN, CAE & PyTorch & Yes & Python \\ 
MesoNet & CNN, CAE & Keras, TensorFlow & No & Python \\  \rowcolor{lavender}
NiftyNet & CNN & TensorFlow & Yes & Python \\ 
ANTsX (ANTsPyNet, ANTsRNet) & CNN, CAE, GAN & Keras, TensorFlow & Yes & Python, R, C++ \\  \rowcolor{lavender}
MARS, BENTO & CNN & TensorFlow & Yes (weights) & Python \\ 
Visual Fields Analysis & DeepLabCut & TensorFlow, DeepLabCut & Yes (weights) & Python \\ 
            
            \bottomrule
        \end{tabularx}
        \captionof{table}{Model engineering specifications for the libraries and frameworks processing image data}
        %\rmfamily
        \label{tab:tabimg2}
%\end{table}

\end{landscape}

\begin{landscape}
%\begin{table}[h]
        \centering
        \small \sffamily
        \begin{tabularx}{600pt}{*{5}{lX} }
            \toprule
            \textbf{Name} & \textbf{Interface} & \textbf{Online/Offline} & \textbf{Maintenance} & \textbf{Source}\\
            \midrule
            \rowcolor{lavender}
            AxonDeepSeg & Jupyter Notebooks & Offline & Active & \url{github.com/axondeepseg/axondeepseg} \\ 
DeepCINAC & GUI, Colab Notebooks & Offline & Active & \url{gitlab.com/cossartlab/deepcinac} \\ \rowcolor{lavender}
DeepLabCut & GUI, Colab Notebooks & Offline & Active & \url{github.com/DeepLabCut/DeepLabCut} \\ 
DeepNeuro & None & Offline & Active & \url{github.com/QTIM-Lab/DeepNeuro} \\ \rowcolor{lavender}
DeepVOG & None & Offline & Inactive & \url{github.com/pydsgz/DeepVOG} \\ 
DeLINEATE & GUI, Colab Notebooks & Offline & Active & \url{bitbucket.org/delineate/delineate} \\ \rowcolor{lavender}
DNNBrain & None & Offline & Active & \url{github.com/BNUCNL/dnnbrain} \\ 
ivadomed & None & Offline & Active & \url{github.com/ivadomed/ivadomed} \\ \rowcolor{lavender}
MEYE & Web app & Online, Offline & Active & \url{pupillometry.it} \\ 
Allen Cell Structure Segmenter & GUI, Jupyter Notebooks & Offline & Active & \url{github.com/AllenCell/aics-ml-segmentation} \\ \rowcolor{lavender}
VesicleSeg & GUI & Offline & Active & \url{github.com/Imbrosci/synaptic-vesicles-detection} \\ 
CDeep3M2 & GUI, Colab Notebooks & Offline & Active & \url{github.com/CRBS/cdeep3m2} \\ \rowcolor{lavender}
CASCADE & GUI, Colab Notebooks & Offline & Active & \url{github.com/HelmchenLabSoftware/Cascade} \\ 
ScLimibic & Unspecified & Offline & Active & \url{surfer.nmr.mgh.harvard.edu/fswiki/ScLimbic} \\ \rowcolor{lavender}
ALMA & GUI & Offline & Active & \url{github.com/sollan/alma} \\ 
fetal-code & GUI, Colab Notebooks & Offline & Active & \url{github.com/saigerutherford/fetal-code} \\ \rowcolor{lavender}
ClinicaDL & GUI, Colab Notebooks & Offline & Active & \url{github.com/aramis-lab/clinicadl} \\ 
DeepNeuron & GUI & Online, Offline & Inactive &   \url{github.com/Vaa3D/Vaa3D_Data/releases/tag/1.0} \\ \rowcolor{lavender}
GaNDLF & GUI & Offline & Active & \url{github.com/CBICA/GaNDLF} \\ 
MesoNet & GUI, Colab Notebooks & Offline & Active & \url{osf.io/svztu} \\ \rowcolor{lavender}
NiftyNet & None & Offline & Inactive & \url{github.com/NifTK/NiftyNet} \\ 
ANTsX (ANTsPyNet, ANTsRNet) & None & Offline & Active & \url{github.com/ANTsX} \\ \rowcolor{lavender}
MARS, BENTO & GUI, MATLAB GUI, Jupyter Notebooks & Offline & Active & \url{github.com/neuroethology} \\ 
Visual Fields Analysis & GUI & Offline & Active & \url{github.com/mathjoss/VisualFieldsAnalysis} \\ 
            \\
            \bottomrule
        \end{tabularx}
        \captionof{table}{Technological aspects and code sources for the libraries and frameworks processing image data}
        %\rmfamily
        \label{tab:tabimg3}
%\end{table}

\end{landscape}
\clearpage
%\twocolumn

\subsection{Libraries targeting data types different from sequences or images and general applications}
%Mnf: Anche supporting libraries non è molto chiaro come titolo. Il titolo qui e sopra devono essere armonizzati. 
%Mnf: Leggendo la prima frase, forse direi "Libraries targeting data types different from sequences or images or general applications". L'ideale sarebbe dividere i due argomenti, ma è da valutare in relazione al numero di tabelle che si possono includere e in relazione al numero di applicazioni.
Libraries and frameworks for sequence data are shown in Tables~\ref{tab:tabxp1} (domains of application),~\ref{tab:tabxp2} (models characteristics),~\ref{tab:tabxp3} (technologies and sources).
In this category fall  libraries and projects with either varying input data type, or other than sequence and image data analysis; other libraries target computational platforms, higher hierarchy frameworks, or supporting functions for deep learning like specific preprocessing and augmentations. 
\textsf{NeuroCAAS} is an ambitious project that both standardizes experimental schedules, analyses and offers computational resources on the cloud. 
The platform lifts the burden of configuring and deploying data analysis tool, guaranteeing also replicability and readily available usage of pre-made pipelines, with high efficiency.
\textsf{MONAI} is a project that brings deep learning tools to many health and biology problems, and is a commonly used framework for the 3D variations of UNet \cite{unet} lately dominating the yearly BraTS challenge \cite{brats} (see at \url{http://braintumorsegmentation.org/}). %
The paradigm builds on PyTorch and aims at unifying healthcare AI practices throughout both academia and enterprise research, not only in the model development but also in the creation of shared annotated datasets. 
Lastly, it focuses on deployment and work in real world clinical production, settling as a strong candidate for being the standard solution in the domain.
\textsf{Predify} and \textsf{THINGvision} are two libraries that bridge deep learning research and computational neuroscience. The former allows to include an implementation of a <<predictive coding mechanism>> (as hypothesized in \cite{predict}) 
into virtually any pre-built architectures, evaluating its impact on performance. 
The latter offers a single environment for Representational Similarity Analysis, i.e. the study of the encodings of biological and artificial neural networks that process visual data.
%%% TorchIO and Envision et al.???

\newpage
\clearpage
\onecolumn
\definecolor{lavender}{rgb}{0.9, 0.9, 0.98}

% &&&&&&&&&&&&&&

\begin{landscape}
        \centering
        
        \small \sffamily
        \begin{tabularx}{600pt}{*{5}{lX} }
            \toprule
            \textbf{Name} & \textbf{Neuroscience area} & \textbf{Data type} & \textbf{Datasets} & \textbf{Task}\\
            
            \midrule
            \rowcolor{lavender}
            NeuroCAAS \cite{noauthor_neuroscience_nodate} & Virtually all & Virtually all & External availability & Virtually all \\ 
MONAI \cite{cardoso_monai_2022} & Virtually all & Virtually all & External availability & Virtually all \\ \rowcolor{lavender}
Predify \cite{choksi_predify_2021} & Computational Neuroscience & Images, Virtually all & No & Classification, Adversarial attacks, virtually all \\ 
THINGvision \cite{muttenthaler_thingsvision_2021} & Computational Neuroscience & Images, Text & External availability & Classification  \\ \rowcolor{lavender}
TorchIO \cite{perez-garcia_torchio_2021} & Imaging & All images & No & Augmentation \\ 
            
            \bottomrule
        \end{tabularx}
        \captionof{table}{Domains of applications for the libraries and frameworks for special applications}
        %\rmfamily
        \label{tab:tabxp1}
\end{landscape}

\begin{landscape}

        \centering
        \small
        \begin{tabularx}{600pt}{*{5}{lX} }
            \toprule
            \textbf{Name} & \textbf{Models} & \textbf{DL framework} & \textbf{Customization} & \textbf{Programming language}\\
            \midrule
            \rowcolor{lavender}
            NeuroCAAS & CNN & TensorFlow & Yes & Python \\ 
MONAI & Virtually All & PyTorch & Yes & Python \\ \rowcolor{lavender}
Predify & CNN, Virtually all & PyTorch & Yes & Python \\ 
THINGvision & CNN, RNN, Transformers & PyTorch, TensorFlow & No & Python \\ \rowcolor{lavender}
TorchIO & CNN & PyTorch & Yes & Python \\ 
            
            \bottomrule
        \end{tabularx}
        \captionof{table}{Model engineering specifications for the libraries and frameworks for special applications}
        %\rmfamily
        \label{tab:tabxp2}

\end{landscape}

\begin{landscape}

        \centering
        \small
        \begin{tabularx}{600pt}{*{5}{lX} }
            \toprule
            \textbf{Name} & \textbf{Interface} & \textbf{Online/Offline} & \textbf{Maintenance} & \textbf{Source}\\
            \midrule
            \rowcolor{lavender}
            NeuroCAAS & GUI, Jupyter Notebooks & Offline & Active & \url{github.com/cunningham-lab/neurocaas} \\ 
MONAI & GUI, Colab Notebooks & Offline & Active & \url{github.com/Project-MONAI/MONAI} \\ \rowcolor{lavender}
Predify & Text UI (TOML) & Offline & Active & \url{github.com/miladmozafari/predify} \\ 
THINGvision & None & Offline & Active & \url{github.com/ViCCo-Group/THINGSvision} \\ \rowcolor{lavender}
TorchIO & GUI, Command line & Offline & Active & \url{torchio.rtfd.io} \\ 
            
            \bottomrule
        \end{tabularx}
        \captionof{table}{Technological aspects and code sources for the libraries and frameworks for special applications}
        %\rmfamily
        \label{tab:tabxp3}

\end{landscape}

\clearpage
%\twocolumn

\section{Discussion}
The panorama of open-source libraries dedicated to deep learning applications in neuroscience is quite rich and diversified.
There is a corpus of organized packages that integrate preprocessing, training, testing and performance analyses of deep neural networks for neurological research.
Most of these projects are tuned to specific data modalities and formats, but some libraries are quite versatile and customizabile, and there are projects that encompass quantitative biology and medical analysis as a whole.
There is a common tendency to develop GUIs, enhancing user-friendliness of toolkits for non-programmers and researchers unacquainted with the command line interfaces, for example. 
Moreover, for the many libraries developed in Python, the (Jupyter) Notebook format appears as a widespread tool both for tutorials, documentation and as an interface to cloud computational resources (e.g. Google Colab \cite{bisong_google_2019}).  
%Mnf: "it is important" è una considerazione soggettiva, quindi meglio evitarla o dire (o aggiungerei) perchè è importante.
Apart from specific papers and documentation, and outside of deep learning per se, it is important to make researchers and developers aware of the main topics and initiatives in open culture and Neuroinformatics. For this reason, the interested reader is invited to rely on competent institutions (e.g. INCF) and databases of open resources (e.g. \textsf{open-neuroscience}) dedicated to Neuroscience. 
Among the possibly missing technologies, the queries employed did not retrieve results in Natural Language Processing libraries dedicated to neuroscience, nor toolkits specifically employing Graph Neural Networks (GNNs), although available in \textsf{EEG-DL}. 
%Mnf: Attenzione: forse menzionerei ad esempio questa: https://link.springer.com/article/10.1007/s12021-018-9404-y
NLP is actually fundamental in healthcare, since medical reports often come in non standardized forms.
Large language models, Named Entity Recognition (NER) systems and text mining approaches in biomedical research exist \cite{DBLP:journals/corr/abs-2007-09134}, \cite{locke_natural_2021}.
%Mnf: qui citerei anche i lavori di ExaMode in NPJ digital medicine e uno sul linguaggio che sta uscendo su Journal of digital pathology
GNNs comprise recent architectures that are extremely promising in a variety of fields \cite{gnnreview}, including biomedical research and particularly neuroscience \cite{zhang_graph_2021}, \cite{li_graph_2022}. 
Even if promising, their application is still less mature than that of computer vision models or time series analysis.

Considering the available software for imaging and signal processing in the domain of neuroscience, at this moment a single alternative targeting the opportunities offered by modern deep learning seems to be missing.
Overall, it seems still unlikely to develop a common deep learning framework for Neuroscience as a separate whole, but the engineering knowledge relevant and compressible into such framework would be common to other biomedical fields, and projects such as \textsf{MONAI} are strong candidates toward this goal. 
Instead, it seems achievable to deliver models and functions in a concerted way, restricted either to a sub-field or a data modality, based on the modularity of existent tools and the organizing possibilities of project initiation and management of open culture.

\section{Conclusions}
Although a large and growing number of repositories offer code to build specific models, as published in experimental papers, these resources seldom aim to constitute proper libraries or frameworks for research or clinical practice.
Both deep learning and neuroscience gain much value even from sophisticated proofs of concept.
In parallel, organized packages are spreading and starting to provide and integrate pre-processing, training, testing and performance analyses of deep neural networks for neurological and biomedical research.
This paper has offered both an historical and a technical context for the use of deep neural networks in Neuroinformatics, focusing on open-source tools that scientists can comprehend and adapt to their necessities.
At the same time, this work underlines the value of the open culture and points to relevant institutions and platforms for neuroscientists.
Although the aim is not restricted to making clinicians develop their own deep models without coding or Machine Learning background, as was the case in \cite{faes_automated_2019}, 
the overall effect of these libraries and sources is to democratize deep learning applications and results, as well as standardizing such complex and varied models, supporting the research community in obtaining proper means to an end, and in envisioning then realizing collectively new projects and tools.

\section*{Acknowledgments}
This work was supported by the "Department of excellence 2018-2022" initiative of the Italian Ministry of education (MIUR) awarded to the Department of Neuroscience - University of Padua.

%Bibliography
\bibliographystyle{unsrt}  
\bibliography{references}  
%\bibliography{biblioontherun}
%\bibliography{dl_libs}

\end{document}